
\def\eg{{\it e.g.,\ }}
\def\ie{{\it i.e.,\ }}
\def\el{{\sl{e--l}} }
\def\ee{{\sl{e--e}} }
\def\Ee{{\sl{E--e}} }
\line{\hfil{{\sl Submitted to Physical Review Letters}}}
\vskip .08 true in
\centerline{\bf Accounting for both electron--lattice and electron--electron}
\vskip 1pt
\centerline{\bf coupling in conjugated polymers: minimum total energy}
\vskip 1pt
\centerline{\bf calculations on the Hubbard--Peierls hamiltonian.}

\vskip .2 true in
\centerline{Giuseppe Rossi}
\vskip .1 true in
\centerline{Ford Research Laboratory, }
\centerline{Ford Motor Company, P.O. Box 2053, Mail Drop 3198,}
\centerline{Dearborn, MI 48121-2053.  }

\vskip 0.2 true in

{\bf Absract.}
Minimum total energy calculations, which account for both electron--lattice
and electron--electron interactions in conjugated polymers are performed for
chains with up to eight carbon atoms. These calculations are motivated in part
by recent experimental results on the spectroscopy of polyenes and conjugated
polymers and shed light on the longstanding question of the relative
importance of electron--lattice vs. electron--electron interactions in
determining the properties of these systems.


\vskip 0.3 true in

     A large amount of experimental evidence~[1] regarding
conjugated polymers can be understood in terms of independent
electron theories that account for electron--lattice \hbox{({\sl{e--l}})}
coupling and $\sigma$--bond compressibility~[2--3]. However, there exists a
considerable body of spectroscopic results, concerning especially
the ordering of excited states~[4--9], which cannot be explained
without invoking electron--electron ({\sl{e--e}}) correlations.
Since such different experimental results are usually rationalized in terms of
models which describe adequately only either the \el or the
\ee interaction, different groups of researchers have been led
to emphasize in these systems the importance of one of the two effects at the
expense of the other.

\medskip
     In this letter the results of a set of minimum total energy calculations
which fully include both interactions are presented. There are several reasons
to pursue this goal. From a theoretical standpoint it is natural to assume that
the transfer integrals depend on the distance between carbon sites and that
there is an energy cost involved in stretching a carbon--carbon bond~[2]. It
is also not surprising to find manifestations of \ee
interactions which are not accounted for by models implying complete screening
such as those of \hbox{ref. [2--3]}. On the experimental side there is a
growing
amount of evidence indicating that the ordering of excited states depends
on the specific polymer and in some instances it appears that different
probing techniques lead to different results in this regard~[4--9].
In particular, recent observations in short thiophene oligomers~[6] and in
poly({\it p--}phenylene--vinylene)~[7] show that in these systems the ordering
of the two lowest excited states is reversed compared to that observed in
polyenes~[4--5]. Within the context of the SSH model~[3] it is natural to
interpret this reversal in terms of lack of ground state degeneracy in the
systems of ref.~[6--7]. This is because, upon excitation of one electron from
the highest occupied to the lowest unoccupied molecular orbital, lack of ground
state degeneracy leads to two separate bipolaron levels as opposed to a pair of
degenerate soliton levels. \Ee repulsion favors the $2\,^1A_g$
over the $1\,^1B_u$ level [10], and the results of ref.~[6--7] suggest that
this latter effect is  not strong enough to overcome the energy difference
between bipolaron levels in these systems. These qualitative considerations
hint to the possibility that important physical effects may be overlooked if
the spectroscopic results are interpreted without fully accounting for
\el interactions.

\medskip
Minimum total energy calculations based on the SSH description were presented
in ref.~[10--11]. For a {\it trans--}polyacetylene chain with $N$ carbon atoms,
the starting point is the hamiltonian for the $\pi$ electron system:
$$
H \; =  \;-\; \sum_{n,s}\; \biggl( t_o - \alpha \bigl( u_{n+1} - u_{n} \bigr)
\biggr) \, \cdot \,
\,\bigl( c_{n+1,s}^{\dagger} c_{n,s} \,+\, c_{n,s}^{\dagger} c_{n+1,s}
\bigr) \; + \;  {K \over {2}} \,\sum_{n} \bigl( u_{n+1} - u_{n} \bigr)^2
\eqno (1)
$$
Here $c_{n,s}^{\dagger}$ and $c_{n,s}$ are creation and annihilation
operators for an electron of spin $s$ on site $n$; $u_{n}$ is the displacement
of the $n$--th ion from its equilibrium position, so that $(u_{n+1} - u_{n})$
is the deviation of the length of the $n$--th bond from its equilibrium length.
The first sum describes hopping with transfer depending linearly on bond
length. The energy associated to $\sigma$--bond compressibility is described
by the second term, $K$ being an elastic spring constant. The hamiltonian
of eq. (1) can be rescaled and rewritten in terms of the dimensionless
coordinates
$ \beta_n  =  \alpha \bigl(u_{n+1} \, - \, u_n \bigr)/t_o $
as
$$
{H \over t_o} \; =  \;-\; \sum_{n,s}\; \bigl( 1 - \beta_n \bigr) \, \cdot \,
\,\bigl( c_{n+1,s}^{\dagger} c_{n,s} \,+\, c_{n,s}^{\dagger} c_{n+1,s}
\bigr) \;+ \; \gamma \,\sum_{n} \beta_n^2
\eqno (2)
$$
where $\gamma = {({K t_o}) / ({2 \alpha^2})}$ accounts for the strength of
the \el coupling
(small $\gamma$ corresponds to strong coupling). A diagonalization of the
first term in the r.h.s. of eq.~(2) gives the single particle electronic
energy levels $\epsilon_{m,s}(\{\beta_n\})$. For a given set of
occupation numbers $\nu_{m,s}$, it is possible to determine the values of the
coordinates $\beta_n$  (\ie of the hopping constants) which
minimize the total energy
$$
E_T (\{ \beta_n \}) \; =
\; \sum_{m,s} \, \nu_{m,s} \, \epsilon_{m,s}(\{\beta_n\}) \;+
\; \gamma \,\sum_{i} \beta_i^2 .
\eqno (3)
$$
for a given $\gamma$.
Here the first sum runs over the possible single particle energy levels; the
second is a sum over the $(N-1)$ bonds. It is through such a minimization
procedure that the models of ref.~[2--3] account for \el
interactions. In the ground state of the half filled system this procedure
leads to Peierls dimerization [10].

\medskip
Within the framework outlined above, the first $^1B_u$ excited state
is obtained by moving one of the two electrons occupying the
$N^{th}$ level to the $(N+1)^{th}$ level. In a long chain (even $N \to \infty$)
the set of bond lengths (\ie the set of values of $\beta_n$) which minimizes
$E_T(\{\beta_n\})$ for this electronic configuration displays two kinks: these
delimit a central portion of chain where the dimerization is inverted [10].
Corresponding to this bond geometry the $N^{th}$ and $(N+1)^{th}$ levels
are degenerate and are found at the center of the Peierls gap.
Similar drastic distortions of the ground state bond geometry with inverted
dimerization in the middle of the chain occur for short chains; however finite
size effects modify the kink bond geometry and break the degeneracy
of $N^{th}$ and $(N+1)^{th}$ levels. In all cases the total energy
corresponding to the optimized bond geometry is substantially smaller than the
energy that the system would have for the same electronic configuration in the
ground state bond geometry. From the Franck-Condon principle it should be
expected that absorption experiments probe the situation where bond lengths are
held to their ground state values while fluorescence experiments probe the
spectrum found by optimizing the bond geometry.

\medskip
Similar considerations hold for higher electronic excited states.
Indeed, for realistic values of $\gamma$, if one allows bond geometry
relaxation, the lowest $^1A_g$ excited state corresponds to moving both
the $N^{th}$ level electrons to the $(N+1)^{th}$ level: in the $N \to \infty$
limit both the total energy and the bond geometry corresponding to this
situation are the same as those of the first $^1B_u$ excited state [12].
However, if the bond geometry of the ground state is kept fixed, the
electronic configuration which gives the lowest excited $^1A_g$ state,
is different: it corresponds to moving one of the $N^{th}$ level electrons to
the $(N+2)^{th}$ level~[10].

\medskip
The rest of this letter is devoted to studying how adding Hubbard terms of
the form
$$
{h \over t_o} \; =
\,+\, v_0 \,\sum_{m}\;  n_{m,\uparrow} \, n_{m,\downarrow}
\,+\, v_1 \,\sum_{m}\;  n_{m} \, n_{m+1}
\eqno (4)
$$
to the hamiltonian of eq. (2) modifies the picture presented above. Here, as
usual $v_0$ and $v_1$
describe on site and nearest neighbor \ee repulsion,
$n_{m}$ is the number operator for electrons on site $m$ and $n_{m,\uparrow}$
($n_{m,\downarrow}$) is the number of spin up (down) electrons.
The distinguishing feature of the treatment presented here resides
in the way the hamiltonian
(sum of (2) and (4)) is dealt with. The fermionic part of the hamiltonian is
diagonalized to give the (many body) energy levels $E_m(\{\beta_n\},v_0,v_1)$.
Then the set of values of the coordinates $\beta_n$ which minimize the total
energy
$$
E_{T,m} (\{ \beta_n \},v_0,v_1) \; =
\; E_m(\{\beta_n\},v_0,v_1) \;+ \; \gamma \,\sum_{i} \beta_i^2
\eqno (5)
$$
associated to the $m^{th}$ (many body) level can be determined for given
values of $\gamma$, $v_0$ and $v_1$. This procedure is the natural extension
of that of ref.~[2--3].
The hopping constants (bond lengths) are not forced into configurations
which cease to be optimal when \ee interactions are turned on.

\medskip
The treatment outlined above differs from the explanations usually
given~[13--15] to rationalize the spectroscopic results  of ref.~[4--9]:
these are based on the results of Pariser--Parr--Pople
quantum chemical calculations, where the hopping constant are forced into
a dimerized configuration fixed from the outset. Within this scheme lattice
relaxations are prevented: \ie neither the hopping constants for the
ground state nor those for the excited states are optimized.
Bond lengths are obtained {\it a posteriori} from $\pi$ bond orders.
Hayden and Mele [16] addressed the issue
of geometry optimization in models including \ee interactions:
using an RG method they did obtain the optimized ground state geometry.
However, they computed the energy of the excited states using the ground
state geometry: this procedure does not account for the \el
effects underlying the soliton physics.

\medskip
The program described above has been implemented numerically within the full
basis set of singlet states for half filled systems with up to eight carbon
atoms [17]. The valence bond basis of ref. [18] for the ${\bf S} = 0$ subspace
is used as a starting point. A symmetric fermion hamiltonian is obtained by
changing to a new basis of singlet states by Gram-Schmidt orthogonalization;
standard algorithms can then be used to find the required eigenvalues~[19].
The set of coordinates $\beta_n$ which minimizes the r.h.s. of eq. (5)
must correspond to bond geometries symmetric with respect to the midbond.
Therefore for a system of N sites minimization of
a function of N/2 independent variables is required: the downhill simplex
method has been used for this purpose~[20].

\medskip
Figure 1 displays examples of results obtained in this way~[21]: it shows the
energy (relative to the ground state energy) of the $1\,^1B_u$ and of the
$2\,^1A_g$ states for two different values of $\gamma$
($\gamma = .9$~[22] and $\gamma = 1.2$), $N = 8$ and $v_1 = 0$. The
energies of $1\,^1B_u$ and $2\,^1A_g$ obtained keeping the ground state bond
geometry fixed are also shown. Note that level crossing~[23] between $1\,^1B_u$
and $2\,^1A_g$ occurs at much lower values of $v_0$ for the optimized excited
state bond geometries than for bond lengths fixed to their ground state values.
Also larger $\gamma$'s, \eg smaller \el interactions, lead to
$1\,^1B_u\,$--$\,2\,^1A_g$ crossings at lower values of $v_0$, at least for
$N \le 8$. Including a nearest
neighbor interaction (non vanishing $v_1$) does not (for reasonable values of
the ratio $v_1/v_0$) change the qualitative features of these results.
In the range $0 \le v_0 \le 5$, the main effect is to increase slightly the
$2\,^1A_g$ energy while that of $1\,^1B_u$ is nearly unchanged. As a result,
the $1\,^1B_u\,$--$\,2\,^1A_g$ crossing occurs at slightly higher values of
$v_0$. Results qualitatively similar to these are found both for $N=6$ and
$N=4$. A detailed description of these and of the other numerical results
summarized in this letter will appear in a forthcoming publication.

\medskip
     It should be noted that three dimensionless parameters ($\gamma$, $v_0$
and $v_1$) completely determine the ratios between the energies of the
electronic states as well as the relative size of the hopping integrals. On the
other hand in order to estimate bond lengths and absolute energies additional
phenomenological constants [10] are needed. To avoid introducing other
parameters, table 1 shows examples of how the hopping constants (rather than
the bond lengths) change as \ee interactions are turned on: note
that large hopping constants correspond to short bonds and {\it vice versa}.
The experimental values of the energies for the $2\,^1A_g$ and $1\,^1B_u$
states are close in polyenes: \ie realistic values of $v_0$ correspond to the
$1\,^1B_u\,$--$\,2\,^1A_g$ crossing region. It is clear from table 1 that
the various types of soliton--like bond geometries (and in particular the
reversed bond alternation in the chain center) survive at these levels of
\ee repulsion.

\medskip
     It seems appropriate at this point to comment on a recent letter by
K\"onig and Stollhoff~[24] which called into question the importance of the
Peierls mechanism in determining the ground state dimerization of
{\it trans}--polyacetylene. The results of table 1 for the half filled ground
state are at variance with the conclusions reached by these authors.
These results show that  (for realistic [22] values of $\gamma$)
\ee interactions have little effect on the ground state
hopping constants and are consistent with a picture where the Peierls
mechanism is the main reason for
dimerization. The results of K\"onig and Stollhoff appear due to their failure
to independently fit their {\it ab initio} results to those obtained from a
semi--empirical hamiltonian which does not include correlations.

\medskip
     In order to model systems where ground state degeneracy is lifted as a
consequence of the local molecular structure, an explicitely biased hopping
term
$$
{H_b \over t_o} \; =  \; {t_b \over t_o}\; \sum_{n,s}\;  (-1)^n \cdot \,\,
\bigl( c_{n+1,s}^{\dagger} c_{n,s} \,+\, c_{n,s}^{\dagger} c_{n+1,s} \bigr)
\eqno (6)
$$
has been added to the hamiltonian (here $t_b$ is a site independent
phenomenological constant). Figure 2 shows numerical results for this situation
when $(t_b/t_o)\,=\,.08$, $\gamma = .9$ and $v_1=0$. As anticipated the
$1\,^1B_u\,$--$\,2\,^1A_g$ crossing occurs for higher values of $v_0$ than
before. Again, analog behavior has been obtained for systems with $N=4$ and
$N=6$ and the distorted ``bipolaron''--like bond geometry survives in presence
of \ee repulsion. Although these results agree with the
qualitative arguments presented at the beginning of this letter, in order to
account quantitatively for the findings of ref.~[6--7] computations on larger
systems as well as more realistic forms of the terms lifting ground state
degeneracy are needed. It should be stressed, in this regard, that the energies
$E_{T,m}$ computed in this letter refer to the semiclassical minima relative
to the optimized bond geometries for the $m^{th}$ many electrons level.
Spectroscopic experiments, on the other hand, probe the various vibronic
levels associated to this electronic state.

\medskip
     The results of figs. 1 and 2 show that $1\,^1B_u\,$--$\,2\,^1A_g$ crossing
occurs for sufficiently high values of $v_0$ even if the bond geometries
are held to their ground state configurations. However, failure to account
for lattice relaxation for the electronic excited states [14] amounts to
ignoring a physical ingredient which is essential in interpreting the
available spectroscopic evidence.

\medskip
  In summary, numerical results from a full many body description of conjugated
chains which includes both \el and \ee effects
have been presented for chains with up to eight carbon sites. These systems are
too small to allow reliable extrapolation to $N \to \infty$ of detailed
numerical results such as those for the energy of the excited states (for given
$\gamma$, $v_0$ and $v_1$) or for the strength of the \ee
repulsion at which $1\,^1B_u\,$--$\,2\,^1A_g$ crossing occurs. However, in view
of the results of ref. [10--11] it is natural to expect that for long chains
the non linear excitations predicted on the basis of the models of ref. [2--3]
will continue to correspond to the relaxed (minimum energy) bond geometries of
the excited states when realistic \ee interactions are turned
on. Moreover, the effect of such interactions on the electronic excitation
spectrum for long chains [12] will be qualitatively similar to that discussed
here for smaller systems.

\medskip
     Several useful discussions with Ken Hass, Phil Pincus and Bill Schneider
are gratefully acknowledged.

\vfill
\eject
\centerline {\bf References and Notes}
\bigskip
\item{[1]}  See for example A.J. Heeger, S. Kivelson, J.R. Schrieffer
and W.P. Su, Rev. Mod. Phys. {\bf 60}, 781 (1988).
\smallskip
\item{[2]}  H.C. Longuet-Higgins and L. Salem,
Proc. Roy. Soc. (London) {\bf A251}, 172 (1959).
\smallskip
\item{[3]}  W.P. Su, J.R. Schrieffer and A.J. Heeger,
Phys. Rev. Lett. {\bf 42}, 1698 (1979).
\smallskip
\item{[4]}  B.S. Hudson, B.E. Kohler and K. Schulten, in {\it Excited States},
edited by E.C. Lim (Academic, New York, 1982) and B.S. Hudson and B.E. Kohler,
Synth. Metals {\bf 9}, 241 (1984).
\smallskip
\item{[5]}  B.E. Kohler, C. Spangler and C. Westerfield,
J. Chem. Phys. {\bf 89}, 5422 (1988).
\smallskip
\item{[6]}  N. Periasamy, R. Danieli, G. Ruani, R. Zamboni and C. Taliani,
Phys. Rev. Lett. {\bf 68}, 919 (1992).
\smallskip
\item{[7]}  C.J. Baker, O.M. Gelsen and D.D.C. Bradley,
Chem. Phys. Lett. {\bf 201}, 127 (1993).
\smallskip
\item{[8]}  B.E. Kohler and D.E. Schilke, J. Chem. Phys. {\bf 86}, 5214 (1987).
\smallskip
\item{[9]}  Y. Tokura, Y. Oowaki, T. Koda and R.H. Baughman,
Chem. Phys. {\bf 88}, 437 (1984).
\smallskip
\item{[10]}  G. Rossi, J. Chem. Phys. {\bf 94}, 4031 (1991).
\smallskip
\item{[11]}  G. Rossi, Synth. Metals {\bf 49}, 221 (1992).
\smallskip
\item{[12]}  If the $N^{th}$ and $(N+1)^{th}$ electronic levels for the first
$^1B_u$ excited state are degenerate there is no additional energy penalty in
moving a second electron from the $N^{th}$ to the $(N+1)^{th}$ level. In a
finite system the degeneracy is broken and $1 ^1B_u$ lies below $2 ^1A_g$.
\smallskip
\item{[13]} Z.G. Soos and S. Ramasesha, Phys. Rev.  {\bf B29}, 5410 (1984).
\smallskip
\item{[14]} Z.G. Soos, S. Etemad, D.S. Galv\~ao and S. Ramasesha,
Chem. Phys. Lett. {\bf 194}, 341 (1992); P.C.M. McWilliams, G.W. Hayden
and Z.G. Soos, Phys. Rev.  {\bf B43}, 9777 (1991);
Z.G. Soos, S. Ramasesha, D.S. Galv\~ao, R.G. Kepler and S. Etemad,
Synth. Metals {\bf 54}, 35 (1993);
Z.G. Soos, S. Ramasesha and D.S. Galv\~ao,
Phys. Rev. Lett. {\bf 71}, 1609 (1993).
\smallskip
\item{[15]} P. Tavan and K. Schulten, Phys. Rev.  {\bf B36}, 4333 (1987).
\smallskip
\item{[16]} G.W. Hayden and E.J. Mele, Phys. Rev. {\bf B34}, 5484 (1986).
\smallskip
\item{[17]} For a half filled system with N sites the size $P$ of the full
basis for the ${\bf S} = 0$ subspace grows exponentially with N:
$$P\,=\, {1 \over {N+1}} {{N+1} \choose {N/2}}^2 \;\;\; .$$
As a result, in order to deal with $N$ significantly larger than eight, one
needs to resort to reduced basis sets (possibly similar to those of
ref. [14--15]).
\smallskip
\item{[18]} S. Mazumdar and Z.G. Soos, Synth. Met. {\bf 1}, 77 (1979).
\smallskip
\item{[19]} For $ N > 6$, if only the lowest few eigenvalues are required,
the Rutishauser algorithm implemented in the NAG Fortran Library -- see
{\it NAG Fortran Library Manual Mark 15}, (NAG Ltd., Oxford, 1991) -- is
considerably more efficient than Householder reduction to tridiagonal form
followed by the QL algorithm: for this procedure see B.T.~Smith, J.M.~Boyle,
B.S.~Garbow, Y.~Ikebe, V.C.~Klema, C.B.~Moler, {\it Matrix Eigensystem
Routines - EISPACK Guide}.  (Springer, Berlin, 1976).
\smallskip
\item{[20]} J.A. Nelder and R. Mead, Comp. Jour. {\bf 7}, 308 (1965).
\smallskip
\item{[21]} Each point in figure 1 involves a few hundred evaluations of
the r.h.s. of eq. (5): \ie diagonalizations of a $P \times P$ matrix
with $P=1764$. A few hours CPU on an HP 9000/735 workstation are required.
\smallskip
\item{[22]}  The choice $\gamma = .9$
has been used to show that minimum energy calculations on the SSH hamiltonian
can reproduce within two percentage points {\it ab initio} results for the bond
lengths of $22$ carbons polyenes [10]. For long chais this value of $\gamma$
yields the accepted kink size in {\it trans--}polyacetylene.
\smallskip
\item{[23]} Further level crossings between $1\,^1B_u$ and higher $^1A_g$
states occur within the range of $v_0$ shown in figs. 1 and 2.
The values of $\beta_n$ which minimize
$E_{T,m}$ change continuously with $v_0$ until a level crossing is reached:
in the crossing region the bond geometries are exchanged.
\smallskip
\item{[24]} G.~K\"onig and G.~Stollhoff,
Phys. Rev. Lett. {\bf 65}, 1239 (1990).
\vfill
\eject

\centerline {\bf Table 1.}
\bigskip
Results for the hopping constants $(1-\beta_n)$ for chains with $N=8$,
$\gamma=.9$ and $v_1=0$. The $n^{th}$ column gives the $n^{th}$ hopping
constant with the outside bond corresponding to $n=1$ and the central bond
to $n=4$. $N_e$ is the number of electrons. The last set of data refers
to a doped chain.
$$\;$$
\settabs 9 \columns
\+&&& $1 - \beta_1$ & $1 - \beta_2$ & $1 - \beta_3$ & $1 - \beta_4$ \cr
\+ ground state  for $N_e = 8$        \cr
\+   $v_0=0.0$ &&&  $2.041$ &  $1.383$ &  $1.974$ &  $1.412$ \cr
\+   $v_0=2.0$ &&&  $2.019$ &  $1.363$ &  $1.961$ &  $1.389$ \cr
\+   $v_0=4.0$ &&&  $1.951$ &  $1.315$ &  $1.911$ &  $1.332$ \cr
\+   $v_0=6.0$ &&&  $1.839$ &  $1.261$ &  $1.813$ &  $1.270$ \cr
\+            \cr
\+ $1\,^1B_u$ state for $N_e = 8$    \cr
\+   $v_0=0.0$ &&&  $1.829$ &  $1.685$ &  $1.503$ &  $1.856$ \cr
\+   $v_0=2.0$ &&&  $1.813$ &  $1.662$ &  $1.524$ &  $1.803$ \cr
\+   $v_0=4.0$ &&&  $1.701$ &  $1.597$ &  $1.479$ &  $1.769$ \cr
\+                                      \cr
\+ $2\,^1A_g$ state for $N_e = 8$    \cr
\+   $v_0=0.0$ &&&  $1.684$ &  $1.847$ &  $1.172$ &  $2.068$ \cr
\+   $v_0=2.0$ &&&  $1.607$ &  $1.815$ &  $1.344$ &  $1.883$ \cr
\+   $v_0=4.0$ &&&  $1.430$ &  $1.742$ &  $1.469$ &  $1.604$ \cr
\+   $v_0=6.0$ &&&  $1.331$ &  $1.653$ &  $1.404$ &  $1.465$ \cr
\+                                      \cr
\+ ground state  for $N_e = 6$        \cr
\+   $v_0=0.0$ &&&  $1.829$ &  $1.685$ &  $1.503$ &  $1.856$ \cr
\+   $v_0=2.0$ &&&  $1.814$ &  $1.662$ &  $1.524$ &  $1.803$ \cr
\+   $v_0=4.0$ &&&  $1.768$ &  $1.628$ &  $1.518$ &  $1.740$ \cr
\+   $v_0=6.0$ &&&  $1.705$ &  $1.592$ &  $1.498$ &  $1.670$ \cr

\vskip 0.8 true in
\centerline{\bf Figure captions.}
\bigskip
FIGURE 1.  Energy $E_{T}$  (in units of $t_o$), relative to the ground state,
of the $1\,^1B_u$ state (empty squares $\gamma = .9$ , filled squares
$\gamma = 1.2$)  and of the $2\,^1A_g$ state (circles $\gamma = .9$ , bullets
$\gamma = 1.2$). Here $v_1=0$ and $N=8$.  The continuous ($\gamma = 1.2$) and
broken ($\gamma=.9$) curves show the corresponding results obtained from the
fixed ground state bond lengths.
\bigskip
FIGURE 2.  Same as in fig. 1 in a system where ground state degeneracy is
broken by a biased hopping term of the form (6), with $(t_b/t_0)\,=\,.08$.
Here $\gamma=.9$, $v_1=0$ and $N=8$.
\bigskip
\bigskip
\bigskip
\parindent=0pt
{\sl PostScript files containing figures are available on request from:}
rossi@scilab.srl.ford.com
\vfill
\eject
\end